\documentstyle[emulateapj,onecolfloat,psfig]{article}   

\begin{document}
\twocolumn[      

\title{Polarization observations in a low synchrotron emission field at 1.4~GHz}

\author{G. Bernardi\altaffilmark{1,2,3}, E.~Carretti\altaffilmark{2},
R.J.~Sault\altaffilmark{3}, M.J.~Kesteven\altaffilmark{3},
S.~Poppi\altaffilmark{4}, S.~Cortiglioni\altaffilmark{2}}

\affil{$^1$Dip. di Astronomia, Universit\`a degli Studi di Bologna, Via Ranzani
1, I-40127, Bologna, Italy\\ 
$^2$C.N.R./I.A.S.F. Bologna, Via Gobetti 101, I-40129 Bologna, Italy\\ 
e-mail: bernardi, carretti, cortiglioni@bo.iasf.cnr.it\\
$^3$ATNF, P.O. BOx 76, EPPING, NSW, 1710, Australia\\ 
e-mail: rsault, michael.kesteven@atnf.csiro.au\\ 
$^4$C.N.R./I.R.A. Bologna, Via Gobetti 101, I-40129 Bologna, Italy\\ 
e-mail: spoppi@ira.bo.cnr.it.}


\begin{abstract}
We present the first observation of the diffuse polarized
synchrotron radiation of a patch ($\sim 3^\circ \times 3^\circ$) in the
BOOMERanG field, one of the areas with the lowest CMB foreground emission. The
work has been carried out with the Australia Telescope Compact
Array at 1.4~GHz with 3.4~arcmin resolution and sensitivity of $\sim 0.18$~mJy
beam$^{-1}$. The mean polarized signal has been found to be
$P_{rms} = \sqrt{(Q_{rms}^2 + U_{rms}^2)} = 11.6 \pm 0.6$~mK, nearly one
order of magnitude below than in the Galactic Plane.

Extrapolations to frequencies of interest for cosmological investigations
suggest that polarized synchrotron 
foreground noise should allow the detection of the CMB Polarization $E$--mode
already at 32~GHz and make us confident that, at 90~GHz, it is accessible with
no relevant foreground contamination. Last but not least, even the $B$--mode
detection for $T/S > 0.01$ is not ruled out in such a low emission patch.  
\end{abstract}

]

\section{Introduction}
The Cosmic Microwave Background Polarization (CMBP) is one of the most promising
field to date for investigating the cosmological parameters and only recently
its detection has been claimed by the DASI (Kovac et al. 2002) and
WMAP\footnote{http://map.gsfc.nasa.gov/} (Kogut et al. 2003) teams. Several
other experiments are planned to measure and characterize it on 
different angular scales (SPOrt\footnote{http://sport.bo.iasf.cnr.it} (see
Carretti et al. 2002a),
PLANCK\footnote{http://astro.estec.esa.nl/SA-general/Projects/Planck/}, B2K2 
(Masi et al. 2002), BaR--SPOrt (Zannoni et al. 2002) and AMiBA (Kesteven et al.
2002) among the others).

Besides the CMBP low emission level (3-4~$\mu$K on sub--degree scales and $< 1\,
\mu$K on larger ones), difficulties in its detection are mainly related to the
presence of foreground noise both from Galactic and extragalactic origin.
Synchrotron polarized radiation is expected to be the main polarized foreground
up to 100~GHz but it is only partially known: only the Galactic plane is well
surveyed and at frequencies lower than 2.7~GHz (Gaensler et al. 2001, Duncan et
al. 1999 and references therein), whereas the only data available at high
latitude (Brouw \& Spoelstra 1976) are widely undersampled.

High Galactic latitudes are particularly relevant for ground--based and
balloon--borne CMBP experiments, whose observations are constrained to limited
patches of the sky, for which low foreground contamination is mandatory. The
DASI team itself points out the absence of polarized foreground observations in
their field 
(Kovac et al. 2002). This scenario emphasizes the importance to investigate the
polarized foreground emission at high Galactic latitudes and calls for new deep
observations of low emission areas aiming at exploring patches suitable
for CMBP investigations. 

In this {\it Letter} we present the first polarization
observations of the BOOMERanG field at RA $= 5^h$ and DEC $=-49^\circ$ (de
Bernardis et al. 2000) with proper sampling and sensitivity. This patch
has been found to be among the lowest dust and synchrotron total
intensity emission areas in the Southern Hemisphere 
and so is very promising
in polarization. It has also been selected as the Southern target for the
BaR--SPOrt experiment (Carretti et al., 2002b). 

We performed our investigation in a $3^\circ \times 3^\circ$ field at 1.4~GHz,
where the signal is expected to be much higher than at the cosmological window
frequencies ($> 30$~GHz). 

Though we must pay attention in extrapolating low frequency data, our results
are encouraging, suggesting that synchrotron diffuse polarized radiation does
not significantly contaminate CMBP measurements at 90~GHz in this
region. CMBP observations seem to be advisable at 32~GHz, even though the
estimated synchrotron signal is closer to the CMBP level.

\section{Observations}

Observations were made with ATCA (Frater, Brooks \& Whiteoak, 1992), an
East--West synthesis interferometer situated near Narrabri (NSW, Australia),
operated by CSIRO--ATNF. Our dataset is a 49 pointing mosaic, which covers
the $3^\circ \times 3^\circ$ region centred at the coordinate RA $=5^h$ and DEC
$=-49^\circ$: the field position was chosen to minimize the presence of point
sources. The observations were carried out in 9 sessions in June 2002, 
for a total effective time of $\sim 72$~h spread over a quite wide range of hour
angles to get a good $u-v$ coverage. The EW214 configuration was used,
which provides useful sensitivity
on scales ranging from $\sim 30$~arcmin down to the
angular resolution of $\sim 3.4$~arcmin. Observations were carried out at the
central frequency of about 1380~MHz using a $2 \times 128$~MHz bandwidth divided in
two sub--bands of 16 channels, each of 8~MHz. Because of interference
contamination, 6 channels (out of 32) were discarded. The system provides the 
four Stokes parameters $I$, $Q$, $U$, $V$. 

The data reduction was carried out using the
MIRIAD package (Sault, Teuben \& Wright 1995). Bandpass and
gain calibrations were performed each session by observing the calibration
source PKS B1934--638, which is assumed to have a flux density of $14.94$~Jy 
at 1.380~GHz (5\% absolute calibration accuracy: see Reynolds 1994, Ott et al.
1994). With this array, we use a conversion factor between 
flux per beam and brightness temperature of 17.4~K/Jy beam$^{-1}$.   

All calibrated visibilities were then inverted to form $I$, $Q$, $U$ and $V$
images using a natural weighting scheme which gives the best signal--to--noise
ratio. Point sources were directly subtracted from $Q$ and $U$ images using
the MIRIAD task MOSSDI (a Clean-based mosaiced image deconvolver - see
Sault, Staveley-Smith \& Brouw 1996) and the residual images were then formed to
obtain the diffuse emission without point source contamination. $I$, $Q$ and $U$
images were jointly deconvolved using the PMOSMEM algorithm (a joint
polarimetric maximum entropy deconvolver for mosaiced observations - see Sault,
Bock \& Duncan 1999), which recovers the large scale structure measured by the
mosaic. The final sensitivity achieved on Stokes $Q$ and $U$ images is $\sim
0.18$~mJy beam$^{-1}$, where the beam size is $\sim3.4$~arcmin. 

\section{RESULTS AND DISCUSSIONS}

\begin{figure*}
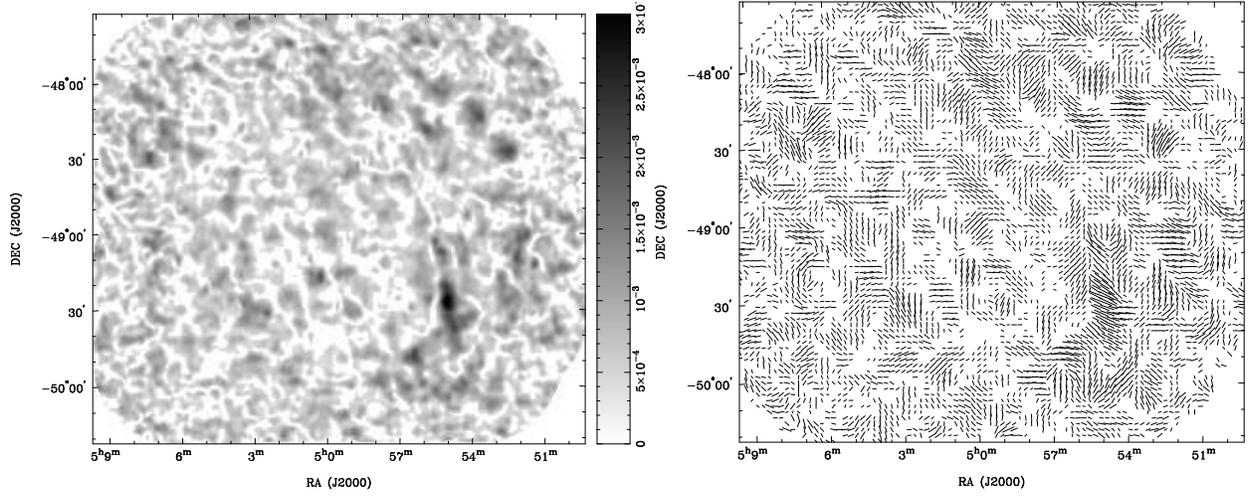
 
 \begin{center}
 \includegraphics[width = 0.35\hsize, angle = -90]{f1a.eps}
 \includegraphics[width = 0.35\hsize, angle = -90]{f1b.eps}
 \end{center}
\caption{Left: map of polarized intensity
$I_p$ (Jy beam$^{-1}$). Pixels with S/N ratio less than $2 \sigma$ are blanked.
Right: polarization angles map. The length of each vector is proportional to the
intensity of the polarized emission at each point. A vector is plotted for every
fourth pixel.}  
\label{pi}
\end{figure*}

\begin{figure*}
 \begin{center}
 \includegraphics[width = 0.35\hsize, angle = -90]{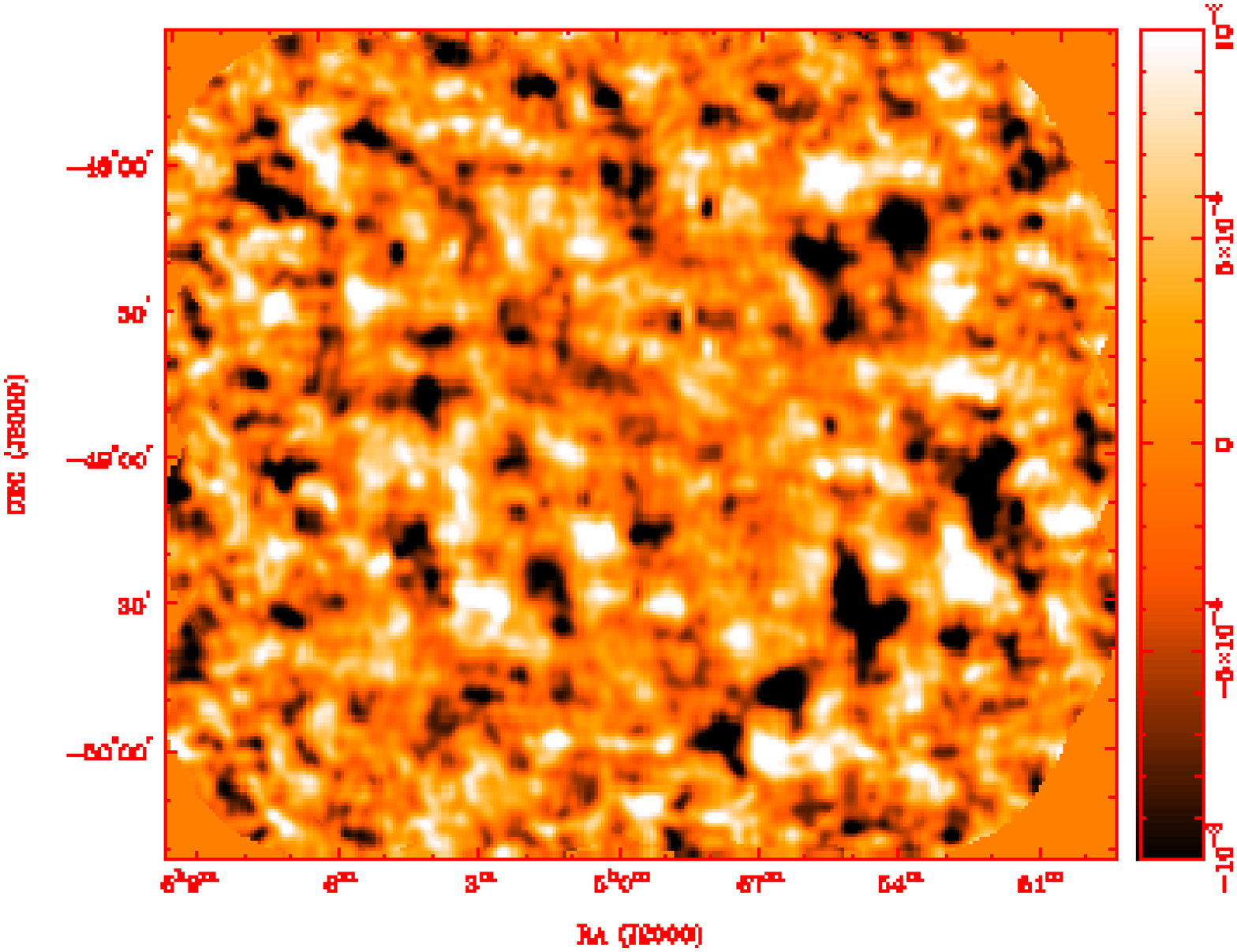}
 \includegraphics[width = 0.35\hsize, angle = -90]{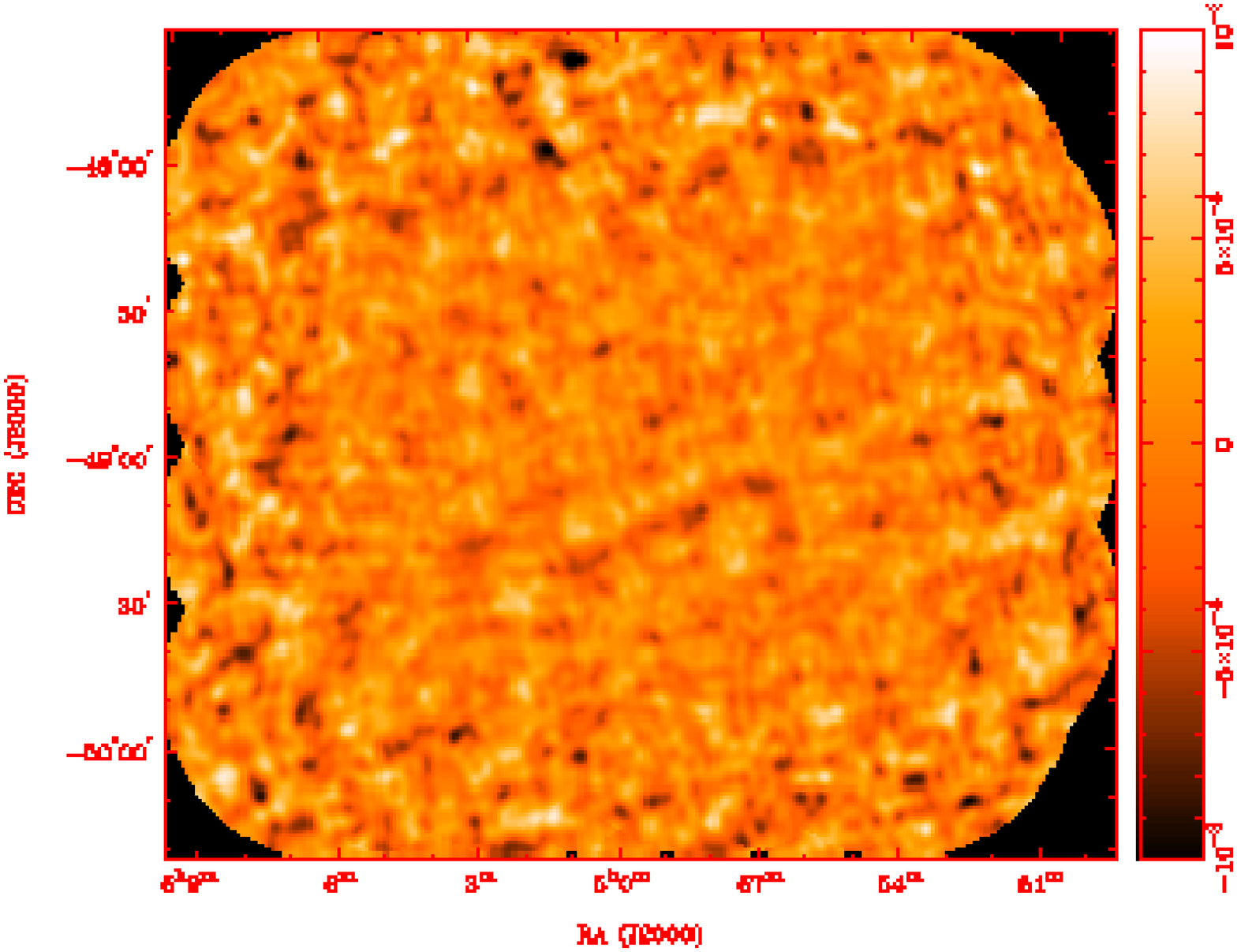}
 \end{center}
\caption{Left: map of the Stokes
parameter $Q$ (Jy beam$^{-1}$). Right: the map of the Stokes parameter $V$ is
shown for comparison (Jy beam$^{-1}$).}
\label{comparison}
\end{figure*}

The polarization maps (intensity and angles) are shown in Figure \ref{pi}.
Stokes parameters $Q$ and $V$ are presented as well (Figure \ref{comparison})
to allow a comparison between signal and noise levels.  
The polarized emission $I_p$ presents extended features distributed over the
whole field. However, a bright feature appears like a filament centred in
RA~$\simeq 4^h55^m$ and DEC~$\simeq -49^\circ 30 \arcmin$ with an extension 
of $\sim 1^\circ$ in declination. This is the brightest polarized structure in
this field, reaching a maximum level of $\sim 3.2$~mJy/beam$^{-1}$. 
The Stokes $I$ image does not show any particular diffuse structure
at the same coordinates even though the strong point source 
contamination makes the comparison difficult. 

It is likely that a {\it Faraday screen} is acting along the line of sight
as a small scale modulation of a relatively uniform background, transfering the
power from larger to smaller scales (for a
comprehensive discussion, see Tucci et al. 2002, and Wieringa et
al. 1993, Gaensler et al. 2001).  Our data shows a uniformity scales for 
polarization angles of 10-15~arcmin and we checked that 
this results in an increasing of
the power of $Q$ and $U$ on the 3-30~arcmin scales to which 
the interferometer is sensitive. 
Thus the detected signal represents an upper limit of the polarized
synchrotron emission in the range where the CMBP peaks ($\sim$~5--30~arcmin).
{\it Faraday screens},
therefore, do not influence the main aim of this work to evaluate the mean
polarized emission, because they produce only a power transfer from large to
small scales. 
 
The Faraday rotation amount is evaluated grouping the 26 useful 
channels in four sub--bands to form four maps at different frequencies. Rotation
Measures (RM) were then deduced from these. RM values were fitted using
the MIRIAD function IMRM, where pixels with an error in the polarization angle
greater than $20^\circ$ were discarded. About $90\%
$ of the obtained RM values are lower than 100 rad/m$^2$ and the RM distribution
peaks at RM~$\sim 50$ rad/m$^2$. These values are consistent with RMs
measurements from extragalactic radio source (Simard--Normandin \& Kronberg
1980) which find values between $|30|$ and $|60|$~rad/m$^2$ at about the same
latitudes. 

The bandwidth depolarization $p$, defined as the
 ratio between the observed and the intrinsic polarization degree, is (Gardner
 \& Whiteoak 1966):   
 \begin{eqnarray}
 p &=& \left | \frac{ \sin (\Delta \Theta)} {\Delta \Theta} \right | \nonumber \\ 
 \Delta \Theta &=& \frac {2 \, \mbox{RM} \, c^2 \, \Delta \nu} {\nu^3}
 \end{eqnarray}
 where $\Delta \nu \sim 205$~MHz is the effective bandwidth and $\nu = 1380$~MHz
 is the central frequency. Our typical RM value of $50$~rad/m$^2$ provides $p
 \sim 0.92$, which is insignificantly different from 1 for our purposes of 
 estimating the mean polarized synchrotron emission in the field. 

 The highest RM values ($\sim 100$~rad/m$^2$) coincides with 
 the brightest polarized structures further supporting their origin from a {\it
 Faraday screen}. 

To estimate the mean polarized synchrotron emission, we computed $rms$ values of
the polarized intensity $I_p$. The $rms$ value of $Q$ and $U$
can be estimated from the maps by removing the noise contribution: 
\begin{eqnarray}
X_{rms} &=& \sqrt{<\tilde{X}^2> - \sigma^2} 
\end{eqnarray}
where $<>$ denotes the average over the area, $\tilde{X}$ the detected signal,
$X$ the intrinsic polarized signal and $\sigma$ the $rms$ noise. The $rms$ value
of $I_p$ is defined as 
\begin{equation}
P_{rms} = \sqrt{Q_{rms}^2 + U_{rms}^2}
\end{equation}
We restricted our analysis to the central $2^\circ \times 2^\circ$ sub--field,
the area  with the highest sensitivity in our field, obtaining 
\begin{eqnarray}
\sigma  &\sim& 3.2 \mbox{mK} \nonumber\\
Q_{rms} &\sim& 8.1 \pm 0.4 \mbox{mK} \nonumber\\
U_{rms} &\sim& 8.4 \pm 0.4 \mbox{mK} \nonumber\\
P_{rms} &\sim& 11.6 \pm 0.6 \mbox{mK} 
\end{eqnarray}
where the error budget is dominated by the calibrator accuracy. If point sources
were not subtracted, we would obtain a slightly greater value, $P_{rms} \simeq
13.0 \pm 0.7$~mK. We note that the polarized intensity mean value is nearly one
order of magnitude lower than the 100--200~mK background emission found near the
Galactic Plane by Uyaniker et al. (1999) at the same 1.4~GHz frequency.

Using the spectral index found by Platania et al. (1998), $\alpha = -3$ in the
1--19~GHz range, $P_{rms}$ can be scaled to estimate the mean polarized signal
at the frequencies of both BaR--SPOrt and B2K2 experiments. Table \ref{scaled}
shows the signal expectation at 32, 90 and 150~GHz as estimated after the
conversion to CMB thermodynamic temperature. We use the conversion factor
\begin{equation}
c  =\left( \frac{2\sinh \frac{x}{2}}{x} \right)^2
\end{equation}
where $x \equiv h \nu/kT_{cmb} \approx \nu/56.8$~GHz.
\begin{deluxetable}{cc} 
\tablecolumns{8} 
\tablewidth{0pc} 
\tablecaption{Estimated foreground noise values at frequencies of cosmological
interest}  
\tablehead{ 
\colhead{$\nu$ (GHz)} & \colhead{Mean polarized signal ($\mu$K)}}
\startdata 
32 &  1.0  \\
90 &  0.05  \\
150 & 0.02  
\enddata 
\label{scaled}
\end{deluxetable}

Since the CMBP signal is expected to be a few $\mu$K on sub--degree angular
scales, both from theoretical predictions (Seljak, 1997) and from 
the recent DASI result (Kovac et al. 2002), our $P_{rms}$ estimate suggests
that, in this patch the polarized synchrotron would not
prevent the detection of the CMBP at a frequency of
32~GHz.

The 90~GHz value is even more encouraging: the estimated synchrotron
contamination is two orders of
magnitude lower than the predicted CMBP signal: it is likely to measure
CMBP without foreground contamination. This conclusion seems to be
unaffected by uncertainties in the synchrotron polarization spectrum: assuming
an uncertainty of $\sim 0.2$ in the spectral index, we obtain a 
worst case foreground signal of $\sim 0.1$~$\mu$K, more than one order of
magnitude lower than CMBP expectations. Considering that the new WMAP results
show a steepening of $\sim 0.5$ in the synchrotron spectral index between K and
Q bands (Bennett et al. 2003), our estimate appears to be conservative,
expecially at 90 and 150~GHz.

Looking at the future, it is worth noting that this patch also 
appears to be
a good candidate for $B$--mode polarization measurements. The
expected $B$--mode level is $P^B_{rms} < 0.3$~$\mu$K for tensor to scalar
perturbation power ratio $T/S < 1$. At this extremely low level, the foreground
contamination should be relevant over most of the sky, jeopardizing
all--sky $B$--mode surveys. Therefore, the detection of $B$--mode polarization
would be possible only in selected regions with low Galactic contamination.

The detection reported in this paper is an upper 
limit of the Galactic synchrotron polarized emission in the 3--30~arcmin 
range and it is not directly comparable with the $B$--mode of the CMBP, which
peaks on 1--2$^\circ$ angular scales.
However, the $B$--mode angular power spectrum of the Galactic synchrotron 
follows a power law behaviour
\begin{equation}
  C_\ell \propto \ell^{\alpha} 
\end{equation}
with 
\begin{equation}
  \alpha > -2.0\,
\end{equation}
and $\ell$ the number of multipole corresponding to the angular scale 
$\theta = 180^\circ / \ell$ (Tucci et al. 2000, Bruscoli et al. 2002).
Thus, the mean polarized emission on a given scale follows a power law
 $P^2_{rms} \propto {\ell (\ell + 1) / (2\, \pi)}\,C_l \, \propto \,\ell^{\gamma}$
with
 $\gamma > 0.0$ ,
providing a polarized emission decreasing with the angular scale. As a result,
the mean polarized emission measured in the 3--30~arcmin range represents an 
upper limit for the $B$--mode emission on the 1-2~degree scale. In turn, this
allows to estimate an upper limit of the Galactic synchrotron contamination 
on the $B$--mode of the CMBP.

The synchrotron emission estimated at 90~GHz in this area ($P_{rms} \sim
0.05$~$\mu$K) corresponds to the $B$--mode signal ($P^B_{rms}$) in models with
$T/S \sim 0.01$, suggesting us the sensitivity for this cosmological parameters 
achievable in this low foreground emission area.

The situation is even better at 150~GHz, where the estimated synchrotron level
corresponds to the $B$--mode emission for $T/S \sim 0.002$. However, at this
frequency, the most important contamination is expected to come from Galactic
dust, for which no data about polarized emission exists in this patch. 
 
\section{CONCLUSIONS}

In this {\it Letter} we present the first polarization measurement of the
BOOMERanG field at 1.4~GHz. 

The polarized emission has been detected and shows a smoothed behaviour with no
peculiar properties. There is, however, a bright structure with no
corresponding feature in total intensity.
The mean polarized emission over
the area has been found to be $P_{rms} = 11.6 \pm 0.6$~mK, nearly one order of
magnitude below the background level in the Galactic Plane.

Our RM estimates suggest that (even though {\it Faraday screens} are 
the likely mechanism)
Faraday depolarization across the bandwidth is negligible, 
allowing a fair estimate of the polarized emission amount.

The extrapolation to higher frequencies suggests that, in this region, $E$--mode
CMBP signal could be measurable even at 32~GHz. At 90~GHz, we
are confident that foreground noise is low enough (more than two order of
magnitude lower than the expected cosmological signal) to not affect the CMBP
signal detection. Last but not least, the low emission at 90~GHz makes this
patch a good candidate also for the $B$--mode detection provided a $T/S > 0.01$.
Even lower $T/S$ values ($\sim 0.002$) could be accessible at 150~GHz, but at
this frequency the signal is more likely affected by dust contamination.

In the light of these results, it is important to make observations at higher
frequencies to confirm such preliminary estimates. Such observations 
would
give us information on the intrinsic properties of the local polarized
foreground, the ISM, and an improved understanding of the role of Faraday
rotation. 

\smallskip
{\it Acknowledgments:} 
We thank Mark Wieringa and the anonymous referee for useful comments. 
This work has been
carried out in the frame of the SPOrt programme funded by the Italian Space
Agency (ASI). G.B. acknowledges a Ph.D. ASI grant. We acknowledge the use of CMBFAST
package. The Australia Telescope Compact Array is part of the Australia
Telescope which is funded by the Commonwealth of Australia for operation as a
National Facility managed by CSIRO.

]

\begin{thebibliography}{}
\bibitem{Be03} Bennett, C.L., et al., 2003, astro--ph/0302208, submitted to ApJ
\bibitem{Br02} Bruscoli, M., Tucci, M., Natale, V., Carretti, E., Fabbri, R.,
	Sbarra, C., Cortiglioni, S., 2002, NewA, 7, 171
\bibitem{Ca02a} Carretti, E., et al., 2002a, in AIP Conf. Proc. 609, Astrophysical
	Polarized Backgrounds, eds. S. Cecchini, S. Cortiglioni, R. Sault \& C.
	Sbarra, (New York: AIP), 109
\bibitem{Ca02b} Carretti, E., et al., 2002b, in AIP Conf. Proc. 616, Experimental
	Cosmology at Millimetre Wavelengths, eds. De Petris M. \& Gervasi M., 
	(New York: AIP), 140
\bibitem{de00} de Bernardis, P., et al., 2000, Nat, 404, 955 
\bibitem{du99} Duncan, A.R., Reich, P., Reich, W., \& F\"urst, E., 1999, A\&A, 350, 447.
\bibitem{fr} Frater, R.H., Brooks, J.W., \& Whiteoak, J.B., 1992, Electrical
	Electron. Eng. Australia, 12 103
\bibitem{ga} Gaensler, B.M., Dickey, J.M., McClure--Griffiths, N.M., Green, A.J.,
	Wieringa, M.H., \& Haynes R.F., 2001, ApJ, 549, 959.
\bibitem{gar} Gardner F.F. \& Whiteoak J.B., 1966, ArA\&A, 4, 245
\bibitem{ke} Kesteven M., 2001, in AIP Conf. Proc. 609, Astrophysical
	Polarized Backgrounds, eds. S. Cecchini, S. Cortiglioni, R. Sault \& C.
	Sbarra, (New York: AIP), 156
\bibitem{kog} Kogut A., et al., 2003, astro-ph/0302213, submitted to ApJ
\bibitem{kov} Kovac, J., Leitch, E.M., Pryke, C., Carlstrom, J.E., Halverson,
	N.W., \& Holzapfel, W.L., 2002, Nature, 420, 772
\bibitem{ma} Masi, S., et al., 2001 in AIP Conf. Proc. 609, Astrophysical
	Polarized Backgrounds, eds. S. Cecchini, S. Cortiglioni, R. Sault \& C.
	Sbarra, (New York: AIP), 122
\bibitem{ot} Ott, M., Witzel, A., Quirrenbach, A., Krichbaum, T.P., Standke, K.J.,
	Schalinski, C.J., \& Hummel, C.J., 1994, A\&A, 284, 331
\bibitem{pl} Platania, P., Bensadoun, M., Bersanelli, M., de Amici, G., Kogut, A.,
	Levin, S., Maino, D. \& Smoot, G.F., 1998, ApJ, 505, 473
\bibitem{re} Reynolds, J.E., 1994, ATNF Technical Document Ser. 39.3040
\bibitem{sa95} Sault, R.J., Teuben, P.J., \& Wright, M.C.H., 1995, in Astronomical
	Data Analysis Software and Systems IV, ed. R. Shaw, H.E. Payne, J.J.E.
	Hayes, ASP, 77, 433 
\bibitem{sa98} Sault, R.J., Bock, D.C.-J., \& Duncan, A.R., 1998, A\&AS, 139, 387
\bibitem{sa96} Sault, R.J., Staveley-Smith, L., \& Brouw, W.N., 1996, A\&AS, 120, 
               375
\bibitem{se} Seljak, U., 1997, ApJ, 482, 6
\bibitem{si} Simard--Normandin, M., \& Kronberg, P.P., 1980, ApJ, 242, 74
\bibitem{tu00} Tucci, M., Carretti, E., Cecchini, S., Fabbri, R., Orsini, M.,
	Pierpaoli, E., 2000, NewA, 5, 181
\bibitem{tu02} Tucci, M., Carretti, E., Cecchini, S., Nicastro, L., Fabbri, R.,
	Gaensler, B.M., Dickey, J.M., \& McClure--Griffiths, N.M., 2002, ApJ,
	579, 607 
\bibitem{uy} Uyaniker, B., F\"urst, E., Reich, W., Reich, P., \& Wielebinski, R.,
	1999, A\&AS, 138, 31
\bibitem{wi} Wieringa, M.H., de Bruyn, A.G., Jansens, D., Brouw, D.N., \& Katgert
	P., 1993, A\&A, 268, 215
\bibitem{za} Zannoni, M., et al., 2001, in AIP Conf. Proc. 609, Astrophysical
	Polarized Backgrounds, eds. S. Cecchini, S. Cortiglioni, R. Sault \& C.
	Sbarra, (New York: AIP), 115
\end{thebibliography}
\end{document}